Intense and Stable Blue Light Emission from $CsPbBr_3$/$Cs_4PbBr_6$ Heterostructures Embedded in Transparent Nanoporous Films.

*Carlos Romero-Pérez[1], Natalia Fernández Delgado[2], Miriam Herrera Collado[2], Mauricio E. Calvo[1*], Hernán Míguez[1*]*

[1]Instituto de Ciencias de Materiales de Sevilla (Consejo Superior de Investigaciones Científicas-Universidad de Sevilla), C/Américo Vespucio, 49, Sevilla, 41092, Spain.

[2] Department of Material Science, Metallurgical Engineering and Inorganic Chemistry IMEYMAT, Facultad de Ciencias (Universidad de Cádiz), Campus Río San Pedro, s/n, Puerto Real, Cádiz, 11510, Spain.

*E-mail: mauricio.calvo@csic.es, h.miguez@csic.es

**Abstract**

Lead halide perovskite nanocrystals are attractive for light emitting devices both as electroluminescent and color converting materials, since they combine intense and narrow emissions with good charge injection and transport properties. However, most perovskite nanocrystals shine at green and red wavelengths, the observation of intense and stable blue emission still being a challenging target. In this work, we report a method to attain intense and enduring blue emission (470-480 nm), with a photoluminescence quantum yield (PLQY) of 40%, originated from very small $CsPbBr_3$ nanocrystals (diameter<3nm) formed by controllably exposing $Cs_4PbBr_6$ to humidity. This process is mediated by the void network of a mesoporous transparent scaffold in which the zero-dimensional (0D) $Cs_4PbBr_6$ lattice is embedded, which allows the fine control over water adsorption and condensation that determines the optimization of the synthetic procedure and, eventually, the nanocrystal size. By temperature dependent photoemission analysis of samples with different [$CsPbBr_3$]/[$Cs_4PbBr_6$] volume ratios, we show that the bright blue emission observed results from the efficient charge transfer to the $CsPbBr_3$ inclusions from the $Cs_4PbBr_6$ host. Our approach provides a means to attain highly efficient transparent blue light emitting films that complete the palette offered by perovskite nanocrystals for lighting and display applications.

**Introduction**

Lead halide perovskites nanocrystals are amid the most studied semiconductor materials for light emitting devices in the last decade,[1] since they fulfill most requirements in terms of tunability of the color emission (through quantum size effects and composition), brightness (with both high absorbance and quantum yield),[2] emission spectral width (close to that of ultrapure emission standards),[3] and ease of processing in the form of films (typically in the liquid phase and at room temperature), as required in the related industry.[4,5] They are used as both electroluminescent[6,7] and color converting layers[8,9] typically emitting in the green and red regions of the electromagnetic spectrum. However, achieving intense and stable blue emission from perovskite nanocrystal solids is still a challenging task.[10] Different strategies have pursued the development of blue light emitting films, being the preferred ones those based on $CsPbX_3$ (with X=Br,I, or mixtures of both) compositions, due to their superior thermal and environmental stability.[11]

Colloidal synthesis allows synthesizing $CsPbBr_3$ quantum dot (QD) suspensions displaying blue emission, as a result of strong quantum size confinement effects, with PLQY near 100%.[12-14] However, processing of these nanocrystals as quantum dot solids, necessary for most practical applications, implies several purification steps followed by deposition on a substrate, which generally cause inhomogeneities and defects due to aggregation, affecting both the optical transparency of the film and lowering the PLQY significantly. Also, many times the observation of blue emission depends on the partial substitution of $Br^-$ or $I^-$ by $Cl^-$, and/or of $Pb^{2+}$ by $Mn^{2+}$, $Cd^2$, $Zn^{2+}$,[15-18] which may favor the generation of deep intra-gap traps,[19] opening non-radiative pathways that lowers the PLQY, or induce photo-segregation.[20] Another approach to achieve blue light emitting films exploits lower dimensional perovskites, such as $CsPb_2Br_5$ and $Cs_4PbBr_6$, which can be considered as 2D and 0D perovskite derivatives. In this strategy, the reduction in the connectivity of the $[PbBr_6]^{4-}$ octahedra network results in larger bandgaps, as a consequence of exciton dissociation derived from weaker coupling between the conduction and valence bands, which favors the observation of blue emission.[21-24] In this approach, the spontaneous growth of 3D perovskite should be prevented to avoid the formation of interphase traps and electron cascades that could promote emission from the lower bandgap 3D phase. In a combined approach, the (spontaneous or induced) growth of perovskite nanocrystals may take place within a lower dimensionality host.[25-29] An example of this is the formation of

CsPbBr$_3$ nanocrystals within a Cs$_4$PbBr$_6$ [30] or CsPb$_2$Br$_5$ [31] matrices, which can result either from an incomplete reaction between the CsBr and PbBr$_2$ precursors or triggered by the presence of water.[32] Although in most cases these CsPbBr$_3$@Cs$_4$PbBr$_6$ heterostructures display green photoemission, there are a few examples of blue emission, although with a significantly lower PLQY than that achieved from green emitters.[33-36]

In recent years, an alternative approach to the synthesis of perovskite nanocrystals has been developed based on a scaffold-assisted route, where the interconnected void space of a mesoporous material acts as a network of nanoreactors,[37] in which lead halide perovskite nanocrystals can be formed.[38-43] This method provides transparent films with PL tunable in a wide spectral range, achievable both by compositional changes and through quantum size effects, along with high PLQY and increased stabilities compared to traditional synthetic routes. In this context, successful synthesis of fully inorganic[44,45] and hybrid-organic[46-48] ABX$_3$ perovskites, with X=Br or I, emitting in the green and red regions of the spectrum has been achieved.

In this work, we demonstrate an alternative synthetic procedure to attain transparent perovskite films displaying blue emission ($\lambda$~478 nm) with PLQY values as high as 40%. In order to do so, we employ transparent mesoporous SiO$_2$ slabs, which act as scaffolds to embed the Cs$_4$PbBr$_6$ precursors and carry out their reaction in a confined environment. Central to the achievement herein reported is the exposure, mediated by the porous scaffold, of the 0D Cs$_4$PbBr$_6$ phase to water vapor, which triggers its gradual and controlled conversion to the nanocrystalline CsPbBr$_3$ phase to yield the final intense blue-light emitting CsPbBr$_3$@Cs$_4$PbBr$_6$@SiO$_2$ heterostructure. By this method, the crystal size is determined by the degree of exposure to humidity of the Cs$_4$PbBr$_6$@SiO$_2$ film, which allows us to achieve an ensemble of very small (<3 nm) CsPbBr$_3$ nanocrystals. Temperature dependent photoluminescence (TDPL) analysis from samples with different [CsPbBr$_3$]/[Cs$_4$PbBr$_6$] ratios provide relevant information on the mechanisms responsible for the highly efficient blue emission observed. We believe this work constitutes a significant step to achieve bright and stable blue color converting layers that can find applications in both lighting and display technologies.

**Results and Discussion**

Encouraged by results reported in the recent studies on $CsPbBr_3@Cs_4PbBr_6$ heterostructures mentioned above, and considering that porous scaffolds provide a means to controllably expose the 0D phase to water by gradual adsorption of the $H_2O$ molecules on the nanopore walls,[45] we performed a two-step synthesis within the void network of a nanoporous film (made of packed 30 nm $SiO_2$ spheres), as described by the following reactions:

$$CsBr + PbBr_2 \rightarrow Cs_4PbBr_6 \qquad (1)$$

$$Cs_4PbBr_6 \xrightarrow{H_2O} CsPbBr_3 + 3\,CsBr \qquad (2)$$

which yielded a $CsPbBr_3@Cs_4PbBr_6@SiO_2$ heterostructure, as only a small portion of the 0D phase transforms into nanocrystalline perovskite. In **Figure 1**a we represent the infiltration and distribution of the precursor solution with CsBr and $PbBr_2$ (molar ratio of 4:1) via spin-coating throughout the scaffold. Embedded $Cs_4PbBr_6$ crystals are formed when the remaining solvent is completely removed during the final annealing step at 100ºC. $SiO_2$ nanoparticles correspond to gray spheres while $Cs_4PbBr_6$ is presented as a light violet covering layer in between $SiO_2$ nanoparticles. Film morphology can be observed in **Figure S1**. X-Ray Diffraction (XRD) characterization, shown in Figure 1b, reveals a pattern consistent with the rhombohedral phase of $Cs_4PbBr_6$ (ICDD:01-073-2478). The peak at $2\theta\cong12.5$ is characteristic of this compound, while the broad band occurring at $15<2\theta<35$ is due to the amorphous $SiO_2$ film. Remarkably, High-Angle Annular Dark-Field Scanning Transmission Electron Microscopy (HAADF-STEM) analysis, obtained from a lamella of the cross-section of the film (Figure 1c), shows the presence of tiny bright spots (i.e., of higher HAADF intensity, as expected for $CsPbBr_3$ phase when compared to $Cs_4PbBr_6$ with sizes comprised between 1 nm and 3 nm, compatible with the expected $CsPbBr_3$ nanocrystals. These are immersed in a light gray compound surrounding the $SiO_2$ particles that can be attributed to $Cs_4PbBr_6$. The crystalline structures hypothesized for each region in the HAADF-STEM image are also drawn in Fig. 1c. Please note that, although it has been widely speculated that $CsPbBr_3$ inclusions form inside the $Cs_4PbBr_6$ continuous lattice.[49-54], direct observation of the perovskite crystallites has remained elusive until now. Figure 1d, schematically shows the 0D/3D system inside a pore section, where the $CsPbBr_3$ inclusions are represented as blue crystallites inserted into the $Cs_4PbBr_6$ continuum, depicted as a thicker indigo colored layer.

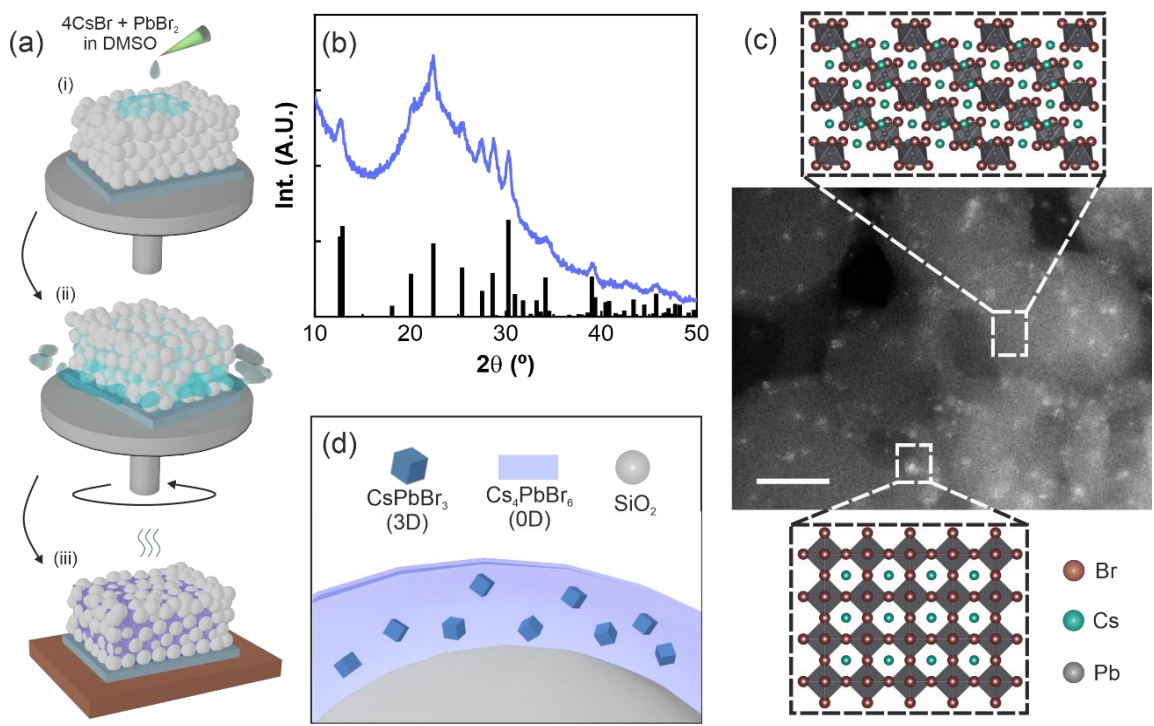

**Figure 1.** (a) Schematic illustration of the synthetic procedure of $Cs_4PbBr_6$ within a $SiO_2$ porous scaffold consisting in (i) solution deposition, (ii) infiltration via spin-coating and (iii) solvent elimination and crystallization through thermal annealing. (b) X-Ray diffractogram of a $Cs_4PbBr_6@SiO_2$ film (indigo line). $Cs_4PbBr_6$ XRD pattern (ICDD:01-073-2478) is represented as black columns. (c) HAADF-STEM micrograph from a $Cs_4PbBr_6@SiO_2$ film lamella. Brighter spots around 1-3 nm size correspond to $CsPbBr_3$ nanocrystals, whose lattice is schematized in the bottom drawing, while the continuum light gray material in which they are embedded is the $Cs_4PbBr_6$ 0D structure, depicted in the top drawing, coating the $SiO_2$ spheres. Scale bar is 10 nm (d) Scheme of the $CsPbBr_3@Cs_4PbBr_6@SiO_2$ film pore section where the $SiO_2$ spheres, the $Cs_4PbBr_6$ layer and the $CsPbBr_3$ nanocrystals are represented in gray, indigo and blue colors, respectively.

The $CsPbBr_3@Cs_4PbBr_6@SiO_2$ films are transparent in the visible region (**Figure S2a**), which is a direct consequence of (i) the use of $SiO_2$ nanoparticles with a size much smaller than visible light wavelength to build the mesoporous scaffold and (ii) the homogeneous infiltration of the heterostructure within its void space, which prevent diffuse light scattering as a result of spatial variations of the refractive index. Also, the $Cs_4PbBr_6$ lattice only absorbs light in the UVB region (Figure S2b), which also contributes to the high average visible transmittance of the films (estimated to be 94.5%, see **Figure S3**). When illuminated under a UV lamp (emission peak at λ≈365 nm), the samples display a strong blue luminescence observable with the naked eye, as shown in **Figure 2a**. For comparison, a picture of a homogeneous bulk $CsPbBr_3@Cs_4PbBr_6$ film, prepared using the same precursors and also exposed to ambient humidity, exhibits a very weak green emission when photoexcited under similar irradiation conditions (also shown in Fig. 2a). The PL spectra of both samples (Figure 2b) confirm that the $CsPbBr_3@Cs_4PbBr_6@SiO_2$ film and the bulk $CsPbBr_3@Cs_4PbBr_6$ film present PL maxima centered in the blue ($\lambda \approx 478$ nm) and the green ($\lambda \approx 525$ nm). Pure $Cs_4PbBr_6$ emits weakly in the near UV region (380-390nm) when exciting at $\lambda=315$ nm, due to the radiative decay of self-trapped excitons originated in the $Pb^{2+}$ states of the $[PbBr_6]^{4-}$ octahedra.[55-57] The visible photoemission observed in moisture exposed $Cs_4PbBr_6$ has been hypothesized to be the consequence of defects localized in the band gap as a result of the presence of bromide vacancies,[58] the formation of polybromide,[59] the presence of interstitial molecules,[60] or the appearance of small quantities of $CsPbBr_3$, be it in the form of inclusions[51,52,54] or as impurities.[49,50] In our case, our hypothesis that the few nanometer size inclusions observed in Fig. 1c are made of $CsPbBr_3$ is supported by the evidence provided by the photoluminescence excitation (PLE) (solid line in Figure 2c) and PL (blue line in Figure 2b) spectra. The PLE, measured fixing the PL collection detector at $\lambda_{em}=478$ nm, is compatible with that of a $CsPbBr_3$ nanocrystals emitting under strong quantum confinement, with an absorption edge at $\approx 450$ nm. Using Brus equation[61] considering a blue-shift of the electronic band, $\Delta E_g$, in the range 450 nm $\leq \lambda \leq$ 480 nm, we can estimate an average $CsPbBr_3$ crystal diameter comprised between 1 and 3 nm, which is consistent with the estimated size of the bright spots observed in the HAADF-STEM micrograph in Figure 1c. The comparison between the PLE and the absorptance (A) spectra of the same $CsPbBr_3@Cs_4PbBr_6@SiO_2$ film (solid and dashed lines in Figure 2c) reveals (i) that the contribution of the $CsPbBr_3$ inclusions to the absorption of the heterostructure is almost imperceptible, contrarily to what is observed in the more sensitive PLE, and (ii) the competition between the absorption processes taking place in the $Cs_4PbBr_6$ host and the 3D inclusions, evidenced by the drop of the PLE intensity from 300 to 330 nm, with a minimum centered at 315 nm, which has been attributed to light absorption in the 0D phase as a result of the formation of excitons in the isolated $[PbBr_6]^{4-}$ octahedra.[56] The absorption band located at higher energies (not fully resolved in Figure 2c) could be assigned to transitions to Pb (6p) from Pb (6s) and Br (4p) orbitals. Excitation wavelength dependence PLQY measurements, whose results are displayed in Fig. 2d show a remarkable 40% quantum efficiency for pump wavelengths at which the only absorption is due to the CsPbBr$_3$ inclusions, one of the highest reported for blue-light emitting perovskite films. It slightly drops to an also significant 30% when both species in the heterostructure are simultaneously excited.

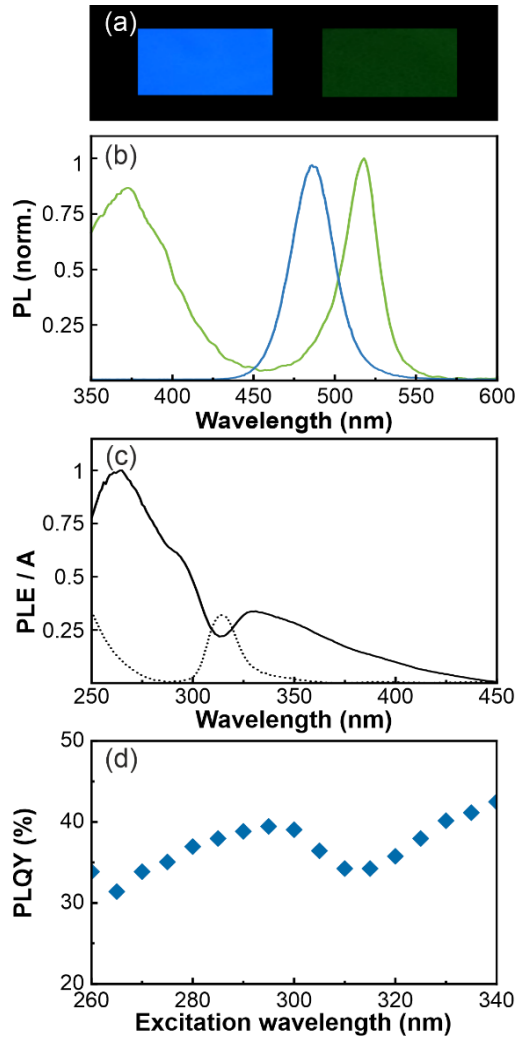

**Figure 2.** (a) Pictures of the samples (left: CsPbBr$_3$@Cs$_4$PbBr$_6$@SiO$_2$, right: CsPbBr$_3$@bulk Cs$_4$PbBr$_6$) attained under UV lamp illumination and (b) PL spectra ($\lambda_{exc}$=315nm) from the same samples (blue and green lines, respectively). (c) PLE, (solid line), A (dotted line) and (d) Excitation wavelength-dependent PLQY of CsPbBr$_3$@Cs$_4$PbBr$_6$@SiO$_2$ films.

As it was mentioned above, the formation of CsPbBr$_3$ inclusions within the Cs$_4$PbBr$_6$ matrix is typically attributed to the exposure of the latter to humidity, and is caused by the highly hygroscopic character of the CsBr present in the 0D phase lattice, which triggers its extraction and hence the change to a less CsBr-rich phase, such as

CsPbBr$_3$.[29,62-64] In our case, the Cs$_4$PbBr$_6$ phase is embedded in a porous scaffold that allows a precise control over the exchange of vapor molecules with the surrounding environment, which is governed by adsorption (and eventually condensation) and desorption processes, characteristic of this kind of mesoporous materials (See **Figure S4**).[65-67] To take advantage of this property, Cs$_4$PbBr$_6$ precursors are infiltrated within the void lattice of the SiO$_2$ scaffold in a nitrogen-filled glovebox with H$_2$O and O$_2$ contents below 0.5 ppm, hence ensuring that no CsPbBr$_3$ is formed in the as-prepared material. Then, without exposing the Cs$_4$PbBr$_6$@SiO$_2$ film to the ambient, it is introduced in a sealed chamber endowed with a quartz window that allows us to perform an in-situ characterization of the PL as the film is gradually exposed to water vapor both in static and dynamic regimes, for which we use the experimental setups depicted in **Figure S5**. A summary of the results obtained are shown in **Figure 3** and commented in detail in what follows.

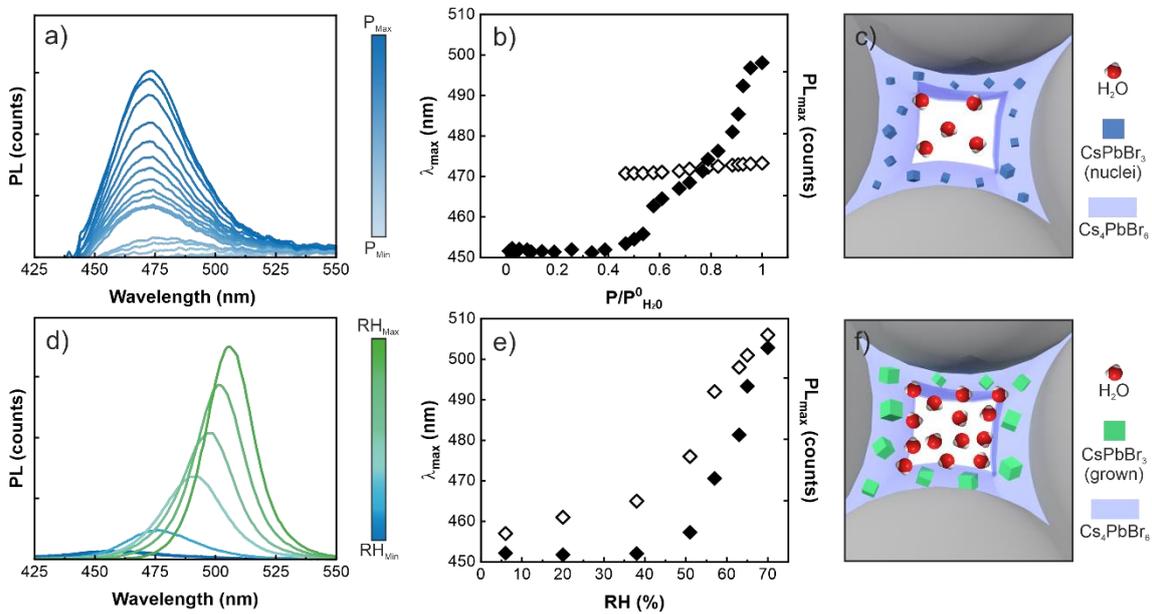

**Figure 3**. Results shown in the upper (a,b,c) and lower (d,e,f) panels correspond to the static and dynamic exposure of the Cs$_4$PbBr$_6$@SiO$_2$ film to water vapor, respectively. (a,d) PL spectra attained at gradually increasing water vapor pressure, (a), or relative humidity, (d), in the chamber. (b,e) PL maximum spectral position (white diamonds) and intensity (black diamonds) of the spectra shown in (a) and (d), respectively. (c,f) Schematic illustration of the hypothesized CsPbBr$_3$ formation within the Cs$_4$PbBr$_6$@SiO$_2$ film in the two water exposure regimes under consideration.

In the static set-up, once the Cs$_4$PbBr$_6$@SiO$_2$ film is transferred from the glovebox (N$_2$ atmosphere) to the sealed chamber (also under N$_2$ atmosphere), the system is vacuum pumped until a pressure of P=0.1 mbar is reached. Then H$_2$O vapor is introduced into the chamber at a preselected pressure (*P*) comprised in the range $0 < \frac{P}{P^0_{H_2O}} < 1$, where $P^0_{H_2O}$ is the water vapor pressure at 298K. In Figure 3a we plot the PL spectra recorded at each $\frac{P}{P^0_{H_2O}}$. Analysis of these curves shows that for $0 < \frac{P}{P^0_{H_2O}} < 0.5$ the PL maximum intensity (*PL$_{max}$*, black diamonds in Figure 3b) is very low, which points at the absence of small CsPbBr$_3$ crystallites throughout the Cs$_4$PbBr$_6$@SiO$_2$ film and indicates that their formation is prevented in the N$_2$-filled glovebox. At $\frac{P}{P^0_{H_2O}} \approx 0.5$, capillary condensation of water in the pores starts, according to Kelvin's law,[65] and a well-defined PL peak with a maximum around $\lambda_{max}$=472nm begins to rise, revealing that the conversion of Cs$_4$PbBr$_6$ to CsPbBr$_3$ in the shape of small crystallites is triggered at that point. Upon increasing pressure, *PL$_{max}$* gradually grows while, interestingly, $\lambda_{max}$ (white diamonds in Figure 3b) remains almost constant, a small 2 nm redshift being detected during the whole process. The Cs$_4$PbBr$_6$ to CsPbBr$_3$ conversion is naturally affected by the diffusion of adsorbed water (the only dominant driving force) within the scaffold, which is determined by the pore size distribution and tortuosity of the SiO$_2$ network. This gives rise to certain broadening of the CsPbBr$_3$ nanocrystal size distribution, which is reflected in the width of the PL spectra (FWHM=37 nm). The subsequent desorption process shows the partial recovery of the originally recorded spectra (see **Figure S6**) indicating that, under these static conditions, the transformation of Cs$_4$PbBr$_6$ to CsPbBr$_3$ is partially reversible.

Interestingly, a very different result is attained when exposure to water vapor occurs in a flow system. In this case, the relative amount of water vapor in the chamber is controlled by pre-mixing different streams of wet and dry nitrogen, setting different relative humidity (RH) values. The PL spectra measured at different RH, as well as the corresponding *PL$_{max}$* (black diamonds) and $\lambda_{max}$ (white diamonds), are plotted in Figures 3d and 3e, respectively. Analogously to the static case, the initial PL is almost undetectable and remains negligible until RH reaches ≈ 50%. Above that value, *PL$_{max}$* starts rising rapidly, while $\lambda_{max}$ undergoes a clearly detectable red-shift, reaching PLQY≈56% (see **Figure S7**) at $\lambda_{max}$ 506 nm for the largest RH achieved (≈70%). Simultaneously, the PL peak width decreases for increasing RH, (FWHM≈23 nm at RH≈70%), pointing at a much narrower CsPbBr$_3$ nanocrystal size distribution than that obtained under static pressure. The gradual shift of the PL peak and its narrowness suggests that the water convection favors the continuous growth of the initially nucleated $CsPbBr_3$ seeds. Although the presence of $CsPbBr_3$ nanocrystals cannot be detected by XRD, which only shows diffraction peaks corresponding to the more abundant $Cs_4PbBr_6$ phase (**Figure S8**), the absorptance of water-exposed $CsPbBr_3@Cs_4PbBr_6@SiO_2$ film increases in the wavelength range 350 nm< $\lambda$ < 510 nm (see **Figure S9**) with respect to the untreated sample (dashed line in Figure 2c), in good agreement with the presence of larger $CsPbBr_3$ nanocrystals, with a red-shifted electronic bandgap. It should also be noted that the conversion of $Cs_4PbBr_6$ into $CsPbBr_3$ is not reversible as in the static system (see **Figure S10**), as $PL_{max}$ only blue-shifts 3 nm upon full desorption of water vapor from the scaffold, indicating that the larger nanocrystals attained in the flow system are more stable than the smaller ones resulting from the static one. Overall, results shown in Fig. 3 reveal that the controlled exposure to water of $Cs_4PbBr_6@SiO_2$, achievable by the intercession of the nanoporous scaffold, allows an unprecedented control over the $CsPbBr_3$ nanocrystal growth process and the final size achieved, rendering a wide chromatic tunability in the blue-green region throughout the process (**Figure S11**).

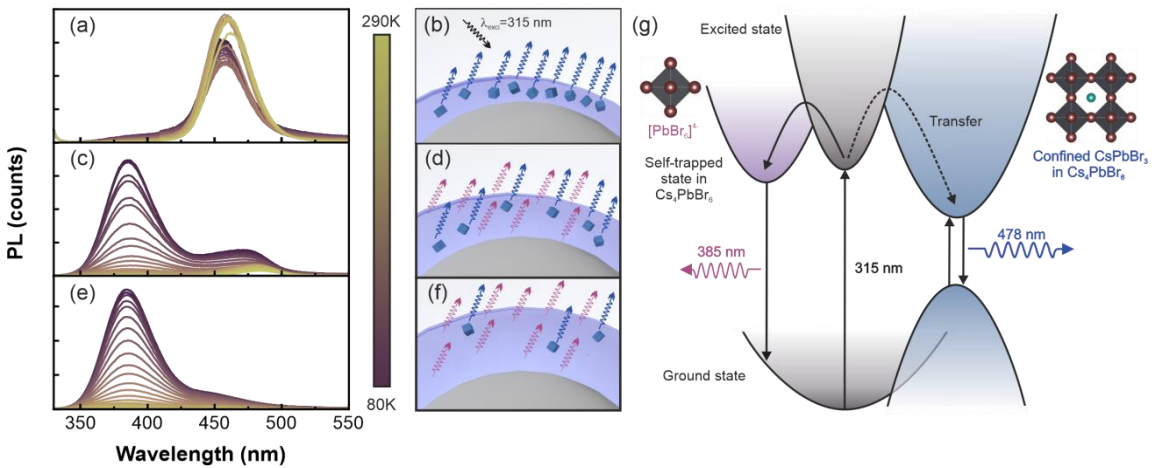

**Figure 4.** Temperature-dependent photoluminescence (left panel) and schematic illustration (central panel) of the spatial distribution of the nanocrystals (blue colored cubes) for gradually increasing coating thickness of $Cs_4PbBr_6$ (violet colored) deposited on the $SiO_2$ pore walls (gray colored). Results correspond to low (a,b), middle (c,d) and high (e,f) $Cs_4PbBr_6$ filling fractions. The color of the PL curves varies from yellow to deep purple as temperature is decreased from 290 K to 80 K inside the cryostat chamber. (g) Simplified energy band diagram and electronic processes leading to emissions originated in a self-trapped state in $Cs_4PbBr_6$ (violet shaded band) and in $CsPbBr_3$ nanocrystals (blue shaded band).

To gain further insight into the mechanisms that generate blue PL in the CsPbBr$_3$@Cs$_4$PbBr$_6$@SiO$_2$ films we performed TDPL measurements in the range 80K<T<290K using $\lambda_{exc}$=315 nm, as both Cs$_4$PbBr$_6$ and CsPbBr$_3$ are photoexcited at that wavelength (see Figure 2c). Results are presented in **Figure 4**. For this analysis, we developed a series of Cs$_4$PbBr$_6$@SiO$_2$ films with three different [CsPbBr$_3$]/[Cs$_4$PbBr$_6$] ratios, which we obtain by varying the concentration of infilled Cs$_4$PbBr$_6$ precursors and hence the thickness of the Cs$_4$PbBr$_6$ coating grown onto the SiO$_2$ pore walls. By doing this, we expect the density of CsPbBr$_3$ nanocrystals embedded within Cs$_4$PbBr$_6$ to significantly decrease with increasing Cs$_4$PbBr$_6$ thickness, since the degree of exposure to water is limited by the coating surface, similar in all three cases. Evidence of the validity of this assumption, as well as more details on the comparative analysis of films with different [CsPbBr$_3$]/[Cs$_4$PbBr$_6$] ratios in terms of their photostability, are provided in the Supporting Information **(Figures S12-S16)**.

Comparison between the results of TDPL of the different concentrations employed allows us to reach some conclusions about the mechanism responsible for the bright blue photoemission observed. First, for the highest [CsPbBr$_3$]/[Cs$_4$PbBr$_6$] ratio (Fig. 4a), we detect, at all temperatures, only a blue PL peak ($\lambda\approx$478nm) corresponding to the emission of CsPbBr$_3$ nanocrystals, with no sign of the ultraviolet emission ($\lambda\approx$385 mn) attributed to the radiative decay of the self-trapped excitons in Cs$_4$PbBr$_6$. However, a very different behavior is observed for the lower [CsPbBr$_3$]/[Cs$_4$PbBr$_6$] ratios (Figs. 4c and 4e), for which photoemission peaks corresponding to both CsPbBr$_3$ and Cs$_4$PbBr$_6$ emissions are clearly detected in the TDPL experiments. Actually, the photoemission of Cs$_4$PbBr$_6$, becomes the prominent emission at some point, outweighing that of CsPbBr$_3$ for T<160K (Figure 4c) and T<260K (Figure 4e), corresponding to the intermediate and minimum [CsPbBr$_3$]/[Cs$_4$PbBr$_6$] ratios, respectively. These results point at a highly efficient electron transfer (as represented in Figure 4g) from the Cs$_4$PbBr$_6$ matrix to the embedded lower bandgap CsPbBr$_3$ nanocrystals in the high [CsPbBr$_3$]/[Cs$_4$PbBr$_6$] ratio samples. The number density of CsPbBr$_3$ nanocrystals decreases in films with lower [CsPbBr$_3$]/[Cs$_4$PbBr$_6$] ratios (as represented in Figures 4b, 4d and 4f), thus lowering the probability to find a CsPbBr$_3$ nanocrystal within the average carrier diffusion distance, measured from an arbitrary photoexcited [PbBr$_6$]$^{4-}$ octahedral site. Hence the PL originated in CsPbBr$_3$ nanocrystals is diminished (Figure 4c) or even almost completely inhibited (Figure 4e), the relaxation of self-trapped excitons in $Cs_4PbBr_6$ becoming the prevalent radiative process in the films. With this analysis in mind, the high brightness observed in Fig. 2a, attained under UV illumination, in spite of the apparent competition for absorbed photons in this spectral range revealed in Fig. 2d, can be explained by the compensating effect of the efficient charge transfer from the host $Cs_4PbBr_6$ matrix to the guest $CsPbBr_3$ inclusions: although the percentage of UV photons absorbed by the film as a whole that decay radiatively in the $CsPbBr_3$ nanocrystals is slightly lower (PLQY≈30%) than when only them are excited ($\lambda_{exc}$>350 nm), the total amount of emitted visible photons is larger because much more photons are absorbed and eventually transferred to the nanocrystals.

**Conclusions**

In essence, we have demonstrated a synthetic route to attain transparent films displaying intense, efficient (40% PLQY) and stable blue light ($\lambda$≈478 nm) emission originated in $CsPbBr_3$ perovskite nanocrystals, formed within a $Cs_4PbBr_6$ matrix previously embedded in a silica mesoporous scaffold. Precise control over the nanocrystal formation process is achieved by gradual exposure of $Cs_4PbBr_6$ to humidity, which is mediated by the adsorption and condensation processes determined by the mesopore network structure along the scaffold. In one approach, our results suggest that forced convection provides a rapid homogenization of the spatial distribution of water vapor within the remaining void space of the $Cs_4PbBr_6$@$SiO_2$ film, accelerating the reaction kinetics and hence allowing a correlation between relative humidity and average crystal size. Finally, the temperature dependent photoemission analysis of samples with different [$CsPbBr_3$]/[$Cs_4PbBr_6$] ratio allows us to demonstrate that the intense blue emission observed results from the efficient charge transfer from the $Cs_4PbBr_6$ host to the $CsPbBr_3$ inclusions. We believe the method herein proposed provides a means to attain highly efficient blue emitting films that can be used in display and lighting technology to complement the color palette offered by perovskite quantum dot color converting layers.

**Experimental Section**

*Materials*

Dimethylsulfoxide (DMSO, Sigma-Aldrich, anhydrous 99.8%), methanol (MeOH, VWR, 98%), 30 nm $SiO_2$ nanoparticles colloidal suspension in $H_2O$ (34% w/v, LUDOX-TMA, Sigma-Aldrich) caesium bromide (CsBr, TCI, 99,9%) and lead (II) bromide ($PbBr_2$, TCI, 99,99%) were used as purchased without conducting any supplementary purification procedure.

*Preparation of $SiO_2$ nanoparticles porous scaffold*

A commercial colloidal suspension of 30 nm in diameter $SiO_2$ nanoparticles was diluted in MeOH to 3% w/v. Dilution was dip-coated on top of a low-fluorescence glass substrate with a 120 mm/min withdrawal speed. Deposition was repeated 15 times in total to produce a porous scaffold around 1.1 μm thick. The scaffold was heated at 450ºC for 30 min to eliminate any organic component remaining within the matrix and to improve mechanical stability.

*Synthesis of $Cs_4PbBr_6$ perovskite inside $SiO_2$ nanoparticles porous scaffold*

A perovskite solution precursor was prepared with CsBr and $PbBr_2$ powders in a 4:1 molar ratio in DMSO at different concentrations (namely low:0.106M, medium: 0.212M and high 0.280M). A spin-coating process was performed at 5000 rpm for 60 seconds to infiltrate the perovskite precursor solution within the porous scaffold. A further annealing procedure at 100ºC for 1 hour was done to remove DMSO as well as to obtain $Cs_4PbBr_6$ crystals inside the porous scaffold. This procedure was done inside a nitrogen-filled glovebox ($H_2O$<0.5ppm; $O_2$<0.5 ppm).

*Characterization*

HAADF-STEM micrographs were obtained from a lamella prepared using a focused ion beam (FIB) instrument (Scios 2 DualBeam, Thermo Fisher Scientific) in combination with a FEI Titan Cubed Themis scanning/transmission electron microscope operated at 200kV.

Grazing incidence X-Ray diffractograms were collected in a Philips X'pert PRO X-Ray diffractometer utilizing Cu cathode Kα radiation (λ=1.54518 Å) in an acquisition range (2θ) of 10°−50° with a 0.05°step.

Total reflectance ($R_t$) and total transmittance ($T_t$) measurements were obtained in a Cary 5000 spectrophotometer (UV−vis−NIR) equipped with an internal DRA-2500 (PMT/PbS version) in the UV-VIS range.

Steady-state PL measurements were analyzed in a spectrometer (Edinburgh FLS1000) with a 450 W ozone-free Xe arc lamp, monochromator in the excitation and the emission and a Red PMT-900 detector. Excitation wavelength is set at different values according to the measurement. Absolute PLQY characterization was done in a C9920-02 Hamamatsu spectrometer equipped with an integrating sphere, a CCD detector and monochromated Xe arc lamp excitation source. Static water exposure PL experiments were done as illustrated in Figure S3a. A sample was attached to a quartz window in a sealed chamber inside a nitrogen-filled glovebox to prevent contact between sample and ambient atmosphere and taken out from the glovebox once prepared and closed. Then it was plugged to rotary vacuum pump and set to vacuum ($5 \times 10^{-2}$ mbar). Water pressure was set opening the purged water vessel line valve and closing it when the desired water pressure was attained and stabilized in the line. The measurement takes place under a 325 nm He–Cd laser as excitation source to collect the PL spectra. This procedure was repeated for different water pressures until the maximum was reached Dynamic measurements were performed according to Figure S3b with sample preparation inside the sealed chamber similar as before. Then it was connected to a system with a dry $N_2$ line and a $N_2$ line passed through a water bubbler. RH was set regulating flows from both streams and measurement was done likewise before. Both experiments were done at 25ºC.

TRPL measurements were performed with 100 ns-pulsed diode laser (454 nm) as excitation source along with a time-correlated single photon counting (TCSPC) detector. TDPL measurements were acquired using a cryostat (Optistat-DM, Oxford Instruments) installed inside the spectrometer. Temperature setting was automatically controlled with the MercuryITC external unit.

**Data availability**

The data that support the findings of this study are openly available in Digital CSIC repository at http://doi.org/[doi] (to be determined in case of final acceptance once an article DOI is provided).

**Conflicts of interest**

There is no conflict of interest to declare.

**Acknowledgements**

This project has received funding from the Spanish Ministry of Science and Innovation under grant PID2020-116593RB-I00, funded by MCIN/AEI/10.13039/501100011033, and of the Junta de Andalucía under grants P18-RT-2291 (FEDER/UE) and PROYEXCEL00955.

Received: ((will be filled in by the editorial staff))
Revised: ((will be filled in by the editorial staff))
Published online: ((will be filled in by the editorial staff))

**Table of contents text**

Gradual exposure to water vapor of $Cs_4PbBr_6$ confined within a nanoporous metal oxide network allows the formation of $CsPbBr_3$ inclusions of controlled size that display intense and efficient (PLQY≈40-50%) blue-to-green emission (λ≈475-515 nm) upon excitation of the $Cs_4PbBr_6$ matrix in the ultraviolet, an effect originated in the host-to-guest transfer of photoexcited carriers, as demonstrated in this work.

**Table of contents figure**

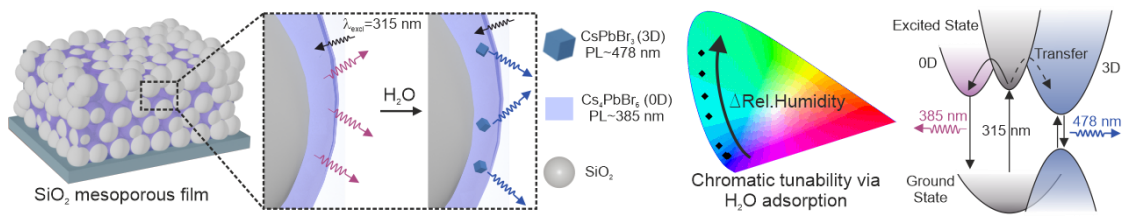